\documentclass[preprint,aps]{revtex4}

\usepackage{graphicx}
\usepackage{dcolumn}
\usepackage{bm}
\usepackage{epsfig}

\begin{document}

\title{Magnification of Subwavelength Field Distributions at Microwave Frequencies
Using a Wire Medium Slab Operating in the Canalization Regime}

\author{Pekka Ikonen$^1$, Pavel Belov$^{2}$, Constantin Simovski$^{1}$, Yang Hao$^2$, and Sergei Tretyakov$^1$}

\affiliation{$^1$Radio Laboratory/SMARAD, TKK Helsinki University of
Technology, P.O.~Box 3000, FI-02015 TKK, Finland\\
$^2$Queen Mary College, University of London, Mile End Road, London, E1 4NS, United Kingdom}

\begin{abstract}
Magnification of subwavelength field distributions using a wire medium slab operating in the canalization regime
is demonstrated using numerical simulations. The magnifying slab is implemented by radially enlarging the
distance between adjacent wires, and the operational frequency is tuned to coincide with the Fabry-Perot
resonance condition. The near-field distribution of a complex-shaped source is canalized over an electrical
distance corresponding roughly to $3\lambda$, and the distribution details are magnified by a factor of three.
The operation of the slab is studied at several frequencies deviating from the Fabry-Perot resonance.
\end{abstract}


\maketitle

Canalization of subwavelength images using electromagnetic crystals was proposed in \cite{Belov1}. Later, two
successful experiments were carried out to demonstrate the canalization of TE-polarized (transverse electric
field with respect to the slab interface) \cite{Ikonen1} and TM-polarized (transverse magnetic field) waves
\cite{Belov_TEM} at microwave frequencies. The slab used in the former experiment was based on capacitively
loaded wires aligned parallel to the slab interfaces, whereas in the latter experiment unloaded wires aligned
perpendicular to the slab interfaces were used. The canalization slab considered in \cite{Belov_TEM} utilizes so
called wire-medium TEM-modes (transverse electromagnetic modes) \cite{Maslovski, spatial} to transport the
details of the source distribution across the slab. Recently, the limitations of subwavelength imaging using
such slabs were analytically studied in \cite{Belov_Mario}, and an experimental study aimed to verification of
the analytical findings is available in \cite{Belov_apl}. Followed by the first studies conducted at microwave
frequencies, authors of \cite{Belov_opt} proposed to implement the canalization regime in the THz range using
stacks of uniaxially positioned alternating dielectric layers.

Recently, motivated mainly by the limitations in the optical microscopy, there has been a growing interest in
structures that are able to \emph{magnify} subwavelength field distributions in the visible range
\cite{Salandrino, Jacob, Liu, Smolyaninov}. This means that the details of the source distribution are retained
while transferring the distribution over a certain distance, and at the same time the distribution is linearly
magnified or enlarged. Essentially, the structures used to demonstrate such an effect in the visible are based
on stacks of two different alternating dielectric layers arranged uniaxially in cartesian or cylindrical
geometries, and one of the layers is implemented as a sheet of plasmonic metal (see also \cite{Belov_opt}). In
the microwave regime Alitalo and co-authors demonstrated experimentally in \cite{Alitalo} simultaneous
enhancement and magnification of evanescent fields (distributions) using double cylindrical polariton-resonant
structures. In this Letter we demonstrate with full-wave simulations the magnification phenomenon at microwave
frequencies using a modified version of the structure considered in \cite{Belov_TEM, Belov_apl}. The proposed
magnifying slab utilizes the canalization phenomenon, thus, it is capable of magnifying distributions comprising
any TM-polarized incident wave (propagating or evanescent) with any transverse component of the wave vector
\cite{Belov1}.


\begin{figure}[t!]
\centering \epsfig{file=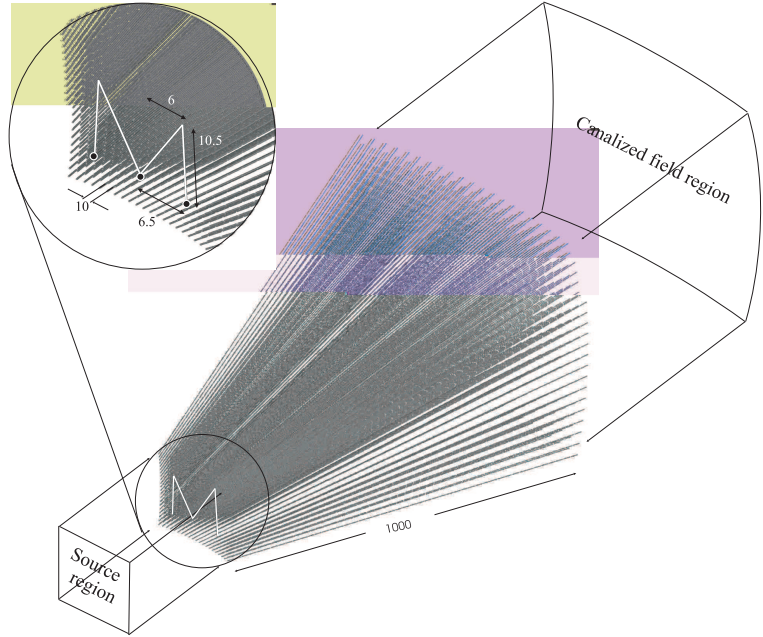, width=10cm} \caption{Schematic illustration of the proposed magnifying slab
(dimensions are in millimeters). The radius of wires $r_0=1$~mm. The white letter ``M'' denotes the wire used as
the source. The feed locations (current sources) are depicted as the black dots in the insert.} \label{sche}
\end{figure}

A schematic illustration of the proposed slab is depicted in Fig.~\ref{sche}. The slab consists of metal wires
(assumed in the simulations to be perfect electric conductors) whose separation is radially enlarged. The wire
ends corresponding to the source interface (input interface of the slab) lie on the surface of a sphere having
radius 500 mm, and the wire ends corresponding to the canalized field interface (output interface) lie on the
surface of a sphere having radius 1500 mm. Altogether $21\times21$ wires comprise the slab, and the separation
between the wire ends at the input interface is approximately $a=10$ mm.

When the operational frequency (with a fixed slab thickness) is tuned to the Fabry-Perot resonance, the
source-field distortion due to reflections is minimized, and the distribution details are transferred across the
slab by the wire-medium TEM-modes. Please note that in theory the Fabry-Perot resonance condition holds for any
(including complex) incidence angle \cite{Belov_TEM}. When the Fabry-Perot resonance condition is met there is
no need, e.g.,~to alter the radius of wires (essentially, to maintain a uniform transmission-line characteristic
impedance), and this significantly eases possible future implementations of the slab. The structure presented in
Fig.~\ref{sche} is expected to yield, in addition to the canalization effect, a magnification of the source
distribution by a factor of three.

The following simulations have been performed using a commercial method-of-moments solver FEKO. The source is a
piece of wire forming letter ``M'' (to reflect ``magnification''), and it is fed by three current sources,
Fig.~\ref{sche}. The distance between the source-wire plane, and the wire end located in the middle of the slab
input interface is 13 mm. The source field distribution is scanned over a planar surface that covers the slab
input interface, and is located at a distance 8 mm behind the source-wire plane. The canalized field
distribution is scanned over a spherical surface that covers the slab output interface, and is located at a 15
mm distance in front of the wire ends. The two scanned regions are in the following referred to as the ``source
region'' and ``canalized field region'', respectively, and they are schematically depicted in Fig.~\ref{sche}.

According to the theory \cite{Belov_TEM, Belov_apl}, only the normal (with respect to the slab interfaces)
component of electric field of a TM-polarized wave is completely (in the ideal case) restored at the slab output
interface. The other two field components contain also contributions from TE-polarized waves, and will in this
implementation be reproduced with distortion. For this reason we present below simulation results only for the
following electric field components:~in the source region we present the component perpendicular to the
source-region plane, and in the canalized field region we present the radial component of electric field.
%

\begin{figure}[t!]
\centering \epsfig{file=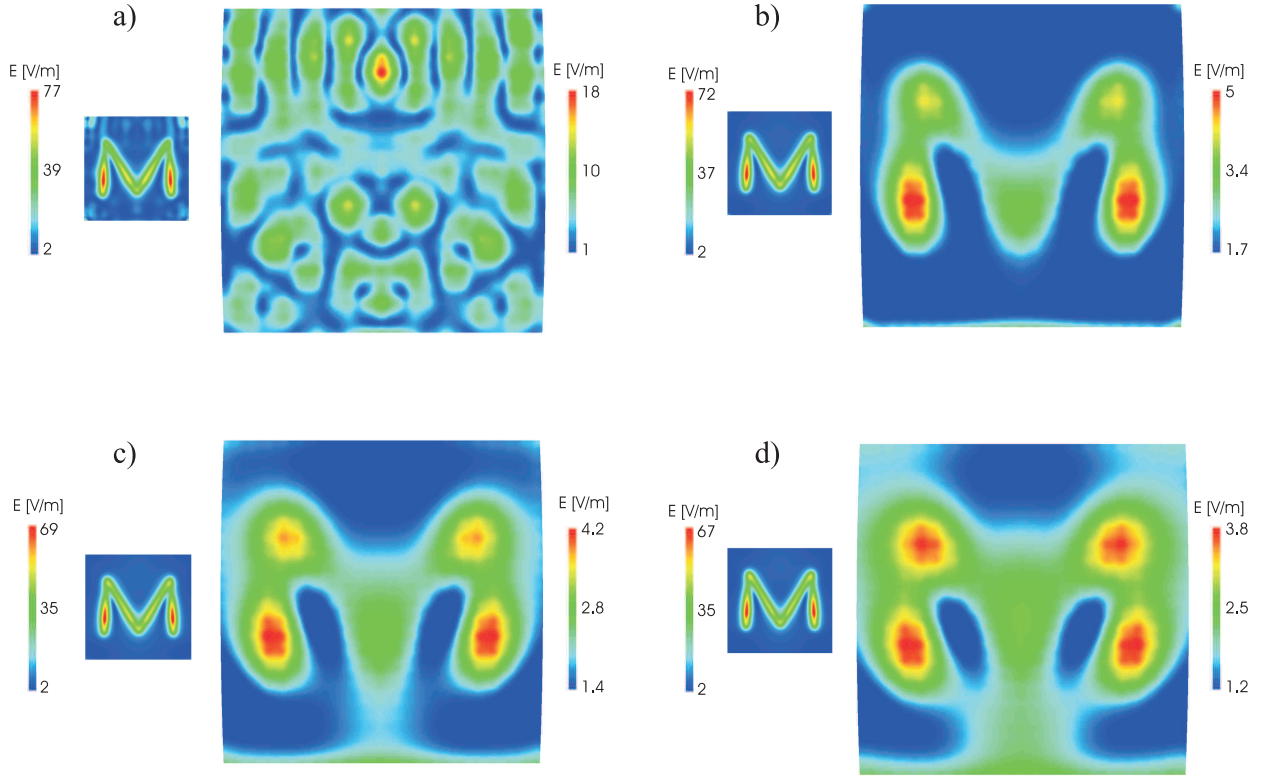, width=16.5cm} \caption{Simulated field distributions at different frequencies.
a) $f=900$ MHz, b) $f=910$ MHz, c) $f=920$ MHz, d) $f=930$ MHz. The relative sizes of the source distribution
and the canalized-field distribution are in scale.} \label{res}
\end{figure}

Fig.~\ref{res} depicts the simulated results. We have performed the simulations at several frequencies in the
vicinity of 900 MHz to identify the frequency that corresponds to the Fabry-Perot resonance (the electrical
thickness of the slab in this frequency range is roughly $3\lambda$). The results indicate that the realized
frequency corresponds roughly to 910 MHz, and at this frequency the electrical length of the wires is
$3.03\lambda$. The small deviation from the theoretical Fabry-Perot condition is most likely caused by the
radially enlarging characteristic dimension of the slab. As predicted, at the realized operational frequency the
source distribution is not affected by reflections, and the details of the distribution are canalized and
simultaneously magnified across the slab. The letter ``M'' (radial electric field component) is accurately
reproduced in the canalized field region, and the characteristic size of the distribution is magnified by a
factor of three. Additional simulations (not shown) indicate that when the canalized field distribution is
scanned very close to the output interface, the re-radiation of the field by the wire ends is clearly visible.
When the canalized field region is located at a distance corresponding to half of the lattice period at the
output interface, this interference vanishes.

Simulations performed at frequencies deviating from the predicted Fabry-Perot resonance indicate the following
(see Fig.~\ref{res}):~as the frequency decreases below the predicted Fabry-Perot resonance, strong interference
appears already after a very small frequency shift. Authors of \cite{Belov_apl} speculated (in connection with
experimental results for the ``regular'' canalization slab) that such interference is mainly caused by strongly
excited surface waves (also the theoretical analysis presented in \cite{Belov_Mario} leads to the same
conclusion). Clearly, the degradation of the canalization effect is not so rapid when the frequency increases
from the predicted Fabry-Perot resonance. The theoretical analysis dealing with regular canalization structures
\cite{Belov_Mario} predicts in this case that the obtainable resolution slightly decreases with increasing
frequency. However, the experimental results presented in \cite{Belov_apl} indicate that such a degradation is
barely visible. Indeed, the meander-line source distribution considered in \cite{Belov_apl} is strongly affected
by reflections from the slab input interface, nevertheless, the modified (distorted) source distribution is
still well canalized at frequencies above the Fabry-Perot resonance. In the simulation scenario considered here
the reflections from the input interface only very moderately affect the details of the source distribution in
the frequency range 910...930 MHz. However, in the canalized field region clear interference caused by the slab
edges is observed as the frequency deviates from the Fabry-Perot resonance. Additional simulations (not shown)
performed in the frequency range 930...1000 MHz indicate that the source-field disturbance caused by the
reflections (with our particular source) is rather moderate over the entire frequency range. The form of the
canalized letter ``M''' can be somehow recognized up to 960 MHz, at higher frequencies the form is
unrecognizable. It is important to observe that the shape, and the technique used to feed the source rather
strongly dictates how noticeable is the degradation of the initial source distribution due to reflections from
the slab interface (this fact is more extensively discussed in \cite{Belov_apl}). Evidently, due to a smaller
amount of subwavelength details, the letter ``M'' is more tolerant to reflections as compared, e.g.,~to the
meander line considered in \cite{Belov_apl}.

In conclusion, in this Letter we have demonstrated simultaneous canalization and magnification of subwavelength
field distributions in the microwave regime using a wire medium slab. The proposed structure consists of an
array of metal wires, and the separation between adjacent wires is radially enlarged. In the example simulations
we have canalized the near field distribution of a complex-shaped source over an electrical distance
corresponding roughly to $3\lambda$, and at the same time magnified the characteristic size of the distribution
by a factor of three. Simulation results showing the operation of the slab at several frequencies deviating from
the operational frequency have been presented and analyzed. In addition to the magnification effect the proposed
slab could be utilized in the opposite way:~electrically large source distributions can be decreased by simply
placing the source in front of the larger slab interface. In this case the source distribution is canalized
across the slab, and the characteristic dimensions of the distribution are simultaneously decreased by a certain
factor.



\end{document}